\documentclass[12pt,cite,epsf,epsfig]{article}
\usepackage{epsfig}

\fontsize{12}{18}

\begin{document}

\author{S. Dev\thanks{%
dev5703@yahoo.com}, Sanjeev Kumar\thanks{%
sanjeev3kumar@yahoo.co.in}, Surender Verma\thanks{%
s\_7verma@yahoo.co.in} and Shivani Gupta\thanks{%
shiroberts\_1980@yahoo.co.in}}
\title{Phenomenological Implications of a Class of Neutrino Mass Matrices}
\date{Department of Physics, Himachal Pradesh University, Shimla 171005, INDIA}

\maketitle

\begin{abstract}
The generic predictions of two-texture zero neutrino mass matrices
of class A in the flavor basis have been reexamined especially in
relation to the degeneracy between mass matrices of types $A_1$
and $A_2$ and interesting constraints on the neutrino parameters
have been obtained. It is shown that the octant of $\theta_{23}$
and the quadrant of the Dirac-type CP-violating phase $\delta$ can
be used to lift this degeneracy.
\end{abstract}

Mass matrices provide important tools for the investigation of the
underlying symmetries and the resulting dynamics. The first step
in this direction is the reconstruction of the neutrino mass
matrix in the flavor basis. However, the reconstruction results in
a large variety of possible structures of mass matrices depending
strongly on the mass scale, mass hierarchy and the Majorana
phases. However, the relatively weak dependence on some
oscillation parameters ($\theta_{23}$ and $\delta$) results in the
degeneracy of possible neutrino mass matrices since all the
parameters are not known at present. The mass matrix for Majorana
neutrinos contains nine physical parameters including the three
mass eigenvalues, three mixing angles and the three CP-violating
phases. The two squared-mass differences ($\Delta m^2_{21}$ and
$\Delta m^2_{32}$) and the two mixing angles ($\theta_{12}$ and
$\theta_{23}$) have been measured in solar, atmospheric and
reactor experiments. The third mixing angle $\theta_{13}$ and the
Dirac-type CP-violating phase $\delta$ are expected to be measured
in the forthcoming neutrino oscillation experiments. The possible
measurement of the effective Majorana mass in neutrinoless double
$\beta$ decay searches will provide an additional constraint on
the remaining three neutrino parameters viz. the neutrino mass
scale and two Majorana CP violating phases. While the neutrino
mass scale will be determined by the direct beta decay searches
and cosmological observations, the two Majorana phases will not be
uniquely determined even if the absolute neutrino mass scale is
known. Under the circumstances, it is natural to employ other
theoretical inputs for the reconstruction of the neutrino mass
matrix. Several proposals have been made in the literature to
restrict the form of the neutrino mass matrix and to reduce the
number of free parameters which include presence of texture zeros
\cite{1,2,3,4}, requirement of zero determinant \cite{5,6} and the
zero trace condition \cite{7} to name a few. However, the current
neutrino oscillation data are consistent only with a limited
number of texture schemes \cite{1,2,3,4}. In particular, the
current neutrino oscillation data disallow all neutrino mass
matrices with three or more texture zeros in the flavor basis. Out
of the fifteen possible neutrino mass matrices with two texture
zeros, only seven are compatible with the current neutrino
oscillation data. The seven allowed two texture zero mass matrices
have been classified into three categories. The two class A
matrices of the types $A_1$ and $A_2$ give hierarchical neutrino
masses. The class B matrices of types $B_1$, $B_2$, $B_3$ and
$B_4$ yield a quasi-degenerate spectrum of neutrino masses. The
single class C matrix corresponds to inverted hierarchy of
neutrino masses. Furthermore, only the mass matrices of class A
and C can accommodate maximal 2-3 mixing  and are, hence, favored
by the data. Thus, only matrices belonging to class A can
accommodate maximal 2-3 mixing and normal hierarchy of neutrino
masses.

In the absence of a significant breakthrough in the theoretical
understanding of the fermion flavors, the phenomenological
approaches are bound to play a crucial role in interpreting new
experimental data on quark and lepton mixing. These approaches are
expected to provide useful hints towards unraveling the dynamics
of fermion mass generation, CP violation and identification of
possible underlying symmetries of the lepton flavors from which
realistic models of lepton mass generation and flavor mixing could
be, hopefully, constructed.

Even though the grand unification on its own does not shed any
light on the flavor problem, the Grand Unified Theories (GUTs)
provide the optimal framework in which possible solutions to the
flavor problem could be embedded. This is because the GUTs predict
definite group theoretical relations between the fermion mass
matrices. For this purpose, it is useful to find out possible
leading order forms of the neutrino mass matrix in a basis in
which the charged lepton mass matrix is diagonal. Such forms of
neutrino mass matrix provide useful hints for model building which
will eventually shed important light on the dynamics of lepton
mass generation and flavor mixing. For example, a
phenomenologically favored texture of quark mass matrix has been
presented earlier \cite{8}. In the spirit of quark-lepton
similarity, the same texture has been prescribed for the charged
lepton and Dirac neutrino mass matrices. The same texture for the
right handed neutrino mass matrix in the see-saw mechanism might
follow from universal flavor symmetry hidden in a more fundamental
theory of mass generation. Thus, the texture zeros in different
positions of the neutrino mass matrix, in particular and fermion
mass matrices, in general could be consequence of an underlying
symmetry. Such universal textures of fermion mass matrices can,
theoretically, be obtained in the context of GUTs based on SO(10).
The four texture zero mass matrices \cite{8} are basically of type
$A_{2}$ in the terminology followed in the present work. Thus, the
selection of $A_{2}$ type neutrino mass matrices will have
important implications for GUTs. On the other hand, there exist
models based on minimal supersymmetric SO(10) \cite{9}, for
example, which predict neutrino mass matrices of type $A_{1}$. An
interesting way of realizing the textures zeros is the triplet
Higgs model based on an abelian discrete symmetry \cite{10} under
which the transformation properties of $\nu_{2}$ and $\nu_{3}$ get
interchanged in neutrino mass matrices of type $A_{1}$ and
$A_{2}$. In another extension of the Standard Model (SM) with
three SU(2) scalar triplets with horizontal symmetry $Z_{6}$
\cite{11}, mass matrices of types $A_{1}$ and $A_{2}$ have
different transformation properties for SU(2) triplets. Thus, the
identification of phenomenologically admissible forms of neutrino
mass matrices will have important implications for the possible
unification schemes at the GUT scale. Furthermore, neutrino mass
matrices of types $A_{1}$ and $A_{2}$ will have, significantly,
different implications for leptonic CP violation which is a
crucial ingredient of leptogenesis to explain the cosmological
baryon asymmetry. The sign of correlation between the Jarlskog
parameter $J$ and the baryon asymmetry is positive for $A_{1}$
type mass matrices while the sign of this correlation could be
both positive or negative for $A_{2}$ type mass matrices leading
to vastly different leptogenesis scenarios \cite{12} for the two
categories of mass matrices.

In the present work, we examine the neutrino mass matrices of
class A for which $m_{ee}=0$ and the neutrino masses are
hierarchical. There are two types of mass matrices in class A
which are consistent with the neutrino oscillation experiments.
For neutrino mass matrices of type $A_1$
\begin{equation}
m_{ee}=m_{e\mu}=0
\end{equation}
whereas
\begin{equation}
m_{ee}=m_{e\tau}=0
\end{equation}
for neutrino mass matrices of type $A_2$. The texture zeros within
a class were thought to have identical phenomenological
consequences \cite{1,2,3} leading to the degeneracy of mass
matrices of types $A_1$ and $A_2$, for example. In the present
work, we discuss the ways to lift this degeneracy. We find that
the knowledge of the deviation of atmospheric mixing from
maximality and of the quadrant of the Dirac-type CP-violating
phase $\delta $ can be used to distinguish the mass matrices of
types $A_1$ and $A_2$. It is, also, found, that the prospects for
the measurement of $\theta_{13}$ in neutrino mass matrices of
class A are quite optimistic since a definite lower bound on
$\theta_{13}$ is obtained for this class.

The neutrino mass matrix, in the flavor basis, is given by
\begin{equation}
(m_{\nu})_{ij}=(Um_{\nu}^{d}U^{T})_{ij};\hspace{10pt}
i,j=e,\mu,\tau \nonumber \\
\end{equation}
for Majorana neutrinos where $m_{\nu}^{d}=Diag\{m_1,m_2,m_3\}$ is
the diagonal neutrino mass matrix and U is the neutrino mixing
matrix which is given by
\begin{equation}
U=\left(
\begin{array}{ccc}
c_{12}c_{13} & s_{12}c_{13} & s_{13}e^{-i\delta } \\
-s_{12}c_{23}-c_{12}s_{23}s_{13}e^{i\delta } &
c_{12}c_{23}-s_{12}s_{23}s_{13}e^{i\delta } & s_{23}c_{13} \\
s_{12}s_{23}-c_{12}c_{23}s_{13}e^{i\delta } &
-c_{12}s_{23}-s_{12}c_{23}s_{13}e^{i\delta } & c_{23}c_{13}
\end{array}
\right) \left(
\begin{array}{ccc}
1 & 0 & 0 \\
0 & e^{i\alpha } & 0 \\
0 & 0 & e^{i\left( \beta +\delta \right) }
\end{array}
\right)
\end{equation}
in PDG representation \cite{13}. Here, $c_{ij}=\cos \theta _{ij}$
and $s_{ij}=\sin \theta _{ij}$. For the above parameterization,
the $ee$, $e \mu$ and $e \tau$ elements of the neutrino mass
matrix are given by
\begin{equation}
m_{ee}=c_{13}^{2}c_{12}^{2}m_{1}+c_{13}^{2}s_{12}^{2}m_{2}e^{2i\alpha
}+s_{13}^{2}m_{3}e^{2i\beta },
\end{equation}
\begin{equation}
m_{e \mu}=c_{13}\{ s_{13}s_{23} e^{i \delta}
( e^{2 i \beta} m_3-s^2_{12} e^{2 i \alpha} m_2 )
-c_{12}c_{23}s_{12} (m_1-e^{2 i \alpha} m_2 )
-c^2_{12} s_{13} s_{23} e^{i \delta} m_1\}
\end{equation}
and
\begin{equation}
m_{e \tau}=c_{13}\{ s_{13}c_{23} e^{i \delta}
( e^{2 i \beta} m_3-s^2_{12} e^{2 i \alpha} m_2 )
+c_{12}s_{23}s_{12} (m_1-e^{2 i \alpha} m_2 )
-c^2_{12} s_{13} c_{23} e^{i \delta} m_1\}.
\end{equation}
It will be helpful to note that $m_{e\tau}$ can be obtained from
$m_{e\mu}$ by exchanging $s_{23}$ with $c_{23}$ and $c_{23}$ with
$-s_{23}$. In other words, the transformation $\theta_{23}
\rightarrow\theta_{23}+\frac{\pi}{2}$ transforms $m_{e\mu}$ to
$m_{e\tau}$. However, the transformation $\theta_{23}\rightarrow
\frac{\pi}{2}-\theta_{23}$ followed by the transformation
$\delta\rightarrow\delta+\pi$, also, transforms $m_{e\mu}$ to
$-m_{e\tau}$. Therefore, if $m_{e\mu}$ vanishes for $\theta_{23}$
and $\delta$, then $m_{e\tau}$ vanishes for
$\frac{\pi}{2}-\theta_{23}$ and $\delta+\pi$. This symmetry
argument clearly shows that the predictions of neutrino mass
matrices for types $A_1$ and $A_2$ will be identical for all
neutrino parameters except $\theta_{23}$ and $\delta$. The
predictions for $\theta_{23}$ and $\delta$ can be parameterized as
\begin{eqnarray}
A_1\hspace{12pt}:\hspace{12pt}\theta_{23}=\frac{\pi}{4}+x,
\hspace{30pt}\delta=y, \nonumber \\
A_2\hspace{12pt}:\hspace{12pt}\theta_{23}=\frac{\pi}{4}-x,
\hspace{12pt}\delta=\pi+y.
\end{eqnarray}

Since, the two squared-mass differences $\Delta m_{12}^{2}$ and
$\Delta m_{23}^{2}$ are known experimentally from the solar,
atmospheric and reactor neutrino experiments, we treat $m_{1}$ as
the free parameter and obtain $m_2$ and $m_3$ from $m_1$ using the
following relations:
\begin{equation}
m_{2}=\sqrt{m_{1}^{2}+\Delta m_{12}^{2}}
\end{equation}
and
\begin{equation}
m_{3}=\sqrt{m_{2}^{2}+\Delta m_{23}^{2}}.
\end{equation}
Our present knowledge \cite{14} of the oscillation parameters has been summarized
below:
\begin{eqnarray}
\Delta m_{12}^{2} &=&7.9_{-0.3,0.8}^{+0.3,1.0}\times 10^{-5}eV^{2},
\nonumber \\
s_{12}^{2} &=&0.31_{-0.03,0.07}^{+0.02,0.09},  \nonumber \\
\Delta m_{23}^{2} &=& 2.2_{-0.27,0.8}^{+0.37,1.1}\times 10^{-3}eV^{2},
\nonumber \\
s_{23}^{2} &=&0.50_{-0.05,0.16}^{+0.06,0.18},  \nonumber \\
s_{13}^{2} &<&0.012(0.046).
\end{eqnarray}
Since, there is only an upper bound on $\theta_{13}$, we treat it
as an unknown parameter in the beginning and use the CHOOZ bound
given above to constrain the allowed parameter space only at the
end. Thus, the element $m_{ee}$ is a function of four unknown
parameters viz. $m_{1}$, $\alpha $, $\beta $ and $\theta _{13}$
while the elements $m_{e\mu}$ and $m_{e\tau}$ are the functions of
five unknown parameters viz.  $m_{1}$, $\alpha $, $\beta $,
$\delta$ and $\theta _{13}$.

The condition $m_{ee}=0$ has, already, been examined in detail
\cite{15}. For $\theta_{13}=0$, $m_{ee}$ vanishes for
$\alpha=90^0$ and $m_1= s^2_{12} \sqrt{\frac{\Delta m^2_{12}}{\cos
2\theta_{12}}}$. For non-zero $\theta_{13}$, the element $m_{ee}$
can vanish only in the normal hierarchy. The real and imaginary
parts of $m_{ee}$ are separately zero if
\begin{equation}
s^2_{13}=\frac{c^2_{12} m_1+s^2_{12} m_2 \cos 2 \alpha}{\mu_1}
\end{equation}
and
\begin{equation}
s^2_{13}=\frac{s^2_{12} m_2 \sin 2 \alpha}{\mu_2}
\end{equation}
where, $\mu_1$ and $\mu_2$ are given by
\begin{equation}
\mu _{1}=m_{1}c_{12}^{2}+m_{2}s_{12}^{2}\cos 2\alpha -m_{3}\cos
2\beta,
\end{equation}
\begin{equation}
\mu _{2}=m_{2}s_{12}^{2}\sin 2\alpha -m_{3}\sin 2\beta.
\end{equation}
The two values of $s^2_{13}$ given in Eq. (12) and Eq. (13) are
equal if
\begin{equation}
\sin 2\beta =-\frac{s_{12}^{2}m_{2}\sin 2\alpha }{M}
\end{equation}
and
\begin{equation}
\cos 2\beta =-\frac{c_{12}^{2}m_{1}+s_{12}^{2}m_{2}\cos 2\alpha }{M}
\end{equation}
and the value of $s^2_{13}$ is given by
\begin{equation}
s_{13}^{2}=\frac{M}{M+m_{3}}
\end{equation}
where
\begin{equation}
M=\sqrt{m_1^2c_{12}^4+m_2^2s_{12}^4+2 m_1 m_2 c_{12}^2 s_{12}^2 \cos 2 \alpha}.
\end{equation}
Using Eqs. (16) and (17), the parameters $\mu_1$ and $\mu_2$ defined in
Eqs. (14) and (15) can be written in terms of a single Majorana
phase $\alpha$ or $\beta$. We express them as functions of
$\beta$:
\begin{eqnarray}
\mu _{1}=-(M+m_3) \cos 2 \beta, \nonumber \\
\mu _{2}=-(M+m_3) \sin 2 \beta.
\end{eqnarray}

Next, we examine the conditions $m_{e\mu}=0$ and $m_{e\tau}=0$.
For vanishing $\theta_{13}$, $m_{e\mu}$ or $m_{e\tau}$ can only be
zero if $m_1=m_2$ and $\alpha=90^0$ which implies a lower bound on
$\theta_{13}$ \cite{4}. Therefore, the angle $\theta_{13}$ will be
bounded from below for neutrino mass matrices of class A.

The element $m_{e\mu}$ vanishes if $Re(m_{e\mu})=0$ and
$Im(m_{e\mu})=0$ which yields the following two relationships for
$s_{13}$:
\begin{equation}
s_{13}=-\frac{c_{12}c_{23}s_{12}}{s_{23}}\frac{m_{1}-m_{2}\cos 2\alpha }{\mu
_{1}\cos \delta -\mu _{2}\sin \delta }
\end{equation}
and
\begin{equation}
s_{13}=\frac{c_{12}c_{23}s_{12}}{s_{23}}\frac{m_{2}\sin 2\alpha }{\mu
_{1}\sin \delta +\mu _{2}\cos \delta }
\end{equation}
where the parameters $\mu_1$ and $\mu_2$ are given by Eqs. (14)
and (15). The two values of $s_{13}$ will be equal if
\begin{eqnarray}
\sin \delta =\frac{\mu _{1}m_{2}\sin 2\alpha -\mu _{2}\left( m_{2}\cos
2\alpha -m_{1}\right) }{\mu_3 \sqrt{\mu_1^2+\mu_2^2}}, \nonumber \\
\cos \delta =\frac{\mu _{2}m_{2}\sin 2\alpha +\mu _{1}\left( m_{2}\cos
2\alpha -m_{1}\right) }{{\mu_3 \sqrt{\mu_1^2+\mu_2^2}}}
\end{eqnarray}
and $s_{13}$ is given by
\begin{equation}
s_{13}=\frac{c_{12}c_{23}s_{12}}{s_{23}}\frac{\mu _{3}}{\sqrt{\mu
_{1}^{2}+\mu _{2}^{2}}}
\end{equation}
where
\begin{equation}
\mu _{3}=\sqrt{m_{1}^{2}+m_{2}^{2}-2m_{1}m_{2}\cos 2\alpha }.
\end{equation}
Therefore, $\delta$ and $s_{13}$ become functions of $m_1$,
$\alpha$ and $\beta$ for vanishing $m_{e\mu}$. Similar expressions
for $\delta$ and $s_{13}$ can be obtained for vanishing $m_{e\tau}$
from Eqs. (23) and (24) by applying the transformations $\theta_{23}
\rightarrow \frac{\pi}{2}-\theta_{23}$ and $\delta\rightarrow \delta
+\pi$. Therefore, $m_{e\tau}$ vanishes if
\begin{eqnarray}
\sin \delta =-\frac{\mu _{1}m_{2}\sin 2\alpha -\mu _{2}\left( m_{2}\cos
2\alpha -m_{1}\right) }{\mu_3 \sqrt{\mu_1^2+\mu_2^2}}, \nonumber \\
\cos \delta =-\frac{\mu _{2}m_{2}\sin 2\alpha +\mu _{1}\left( m_{2}\cos
2\alpha -m_{1}\right) }{{\mu_3 \sqrt{\mu_1^2+\mu_2^2}}}
\end{eqnarray}
and $s_{13}$ is, now, given by
\begin{equation}
s_{13}=\frac{c_{12}s_{23}s_{12}}{c_{23}}\frac{\mu _{3}}{\sqrt{\mu
_{1}^{2}+\mu _{2}^{2}}}.
\end{equation}
The transformation $\delta\rightarrow \delta+\pi$ keeps $\tan
\delta$ unchanged. Equations (26) and (27) can, also, be derived
directly in the same way as we derived Eqs. (23) and (24).

Now, we examine the implications of vanishing $m_{ee}$ and
$m_{e\mu}$ for the neutrino mass matrices of type $A_1$. Since,
$\theta_{13}$ must vanish for $\alpha=90^0$ if $m_{ee}=0$, and a
vanishing $\theta_{13}$ is not allowed by the condition
$m_{e\mu}=0$, the points $\alpha=90^0$ and $\theta_{13}=0$ are not
allowed for the neutrino mass matrices of type $A_1$.

Substituting the values of $\mu_1$ and $\mu_2$ from Eqs. (20) in
Eqs. (23), we obtain
\begin{eqnarray}
\sin \delta =-\frac{m_2 \sin 2 (\alpha-\beta)+m_1 \sin 2 \beta}{\mu_3}, \nonumber \\
\cos \delta =-\frac{m_2 \cos 2 (\alpha-\beta)-m_1 \cos 2
\beta}{\mu_3}.
\end{eqnarray}
Eliminating $\beta$ from the Eqs. (28) by using the Eqs. (16) and (17), we
obtain
\begin{equation}
\sin \delta =\frac{m_1 m_2 \sin 2 \alpha}{M \mu_3}
\end{equation}
and
\begin{equation}
\cos \delta =\frac{s^2_{12} m_2^2-c^2_{12}m_1^2+m_1 m_2
(c^2_{12}-s^2_{12}) \cos 2 \alpha}{M\mu_3}.
\end{equation}
The Dirac-type CP-violating phase $\delta$ can, also, be written in terms
of $\beta$ as
\begin{equation}
\sin \delta =-\frac{m_1 \sin 2 \beta}{s^2_{12} \mu_3}.
\end{equation}
The value $\alpha=90^0$ corresponds to $\beta=0^0$ and
$\delta=0^0$. Since, $\alpha=90^0$ is not allowed, $\beta=0^0$ and
$\delta=0^0$ are, also, not allowed. Therefore, two texture zero
mass matrices of class A are necessarily CP violating. From Eq.
(24), we obtain
\begin{equation}
s_{13}=\frac{c_{12}c_{23}s_{12}}{s_{23}}\frac{\mu_3}{M+\mu_3}.
\end{equation}
For the simultaneous existence of texture zeros at the $ee$ and
$\mu\mu$ entries, the two relations for $s_{13}$ calculated from
the conditions $m_{ee}=0$ and $m_{e\mu}=0$ should be consistent
with each other. Equations (18) and (32) give the same value of
$s_{13}$ for
\begin{equation}
M(M+m_3)\tan^2\theta_{23}=\mu_3^2 s^2_{12}c^2_{12}
\end{equation}
and the value of $s_{13}$ is given by
\begin{equation}
s_{13}=\frac{s^2_{23}}{c^2_{12}c^2_{23}s^2_{12}}\frac{M}{\mu_{3}}.
\end{equation}
Eq. (33) can be solved to obtain $\alpha$ as a function of
$m_1$ for the neutrino mass matrices of type $A_1$. With this
value of $\alpha$, one can calculate
$\delta$ and $\theta_{13}$ from Eqs. (29), (30) and (34).
Eq. (34) can be used to obtain a lower bound on $\theta_{13}$:
\begin{equation}
s_{13}>\frac{s^2_{23}}{c^2_{12}c^2_{23}s^2_{12}}\frac{|m_2 s^2_{12}
-m_1 c^2_{12}|}{m_1+m_2}.
\end{equation}

Similarly, one can show that for neutrino mass matrices of type
$A_2$, $\delta$ and $s_{13}$ are given by
\begin{equation}
\sin \delta =\frac{m_1 \sin 2 \beta}{s^2_{12} \mu_3}
\end{equation}
and
\begin{equation}
s_{13}=\frac{c^2_{23}}{c^2_{12}s^2_{23}s^2_{12}}\frac{M}{\mu_{3}}.
\end{equation}
and the condition for the simultaneous existence of two texture zeros
is
\begin{equation}
M(M+m_3)=\mu_3^2 s^2_{12}c^2_{12}\tan^2\theta_{23}.
\end{equation}
We note the reciprocity between the values of $\tan^2 \theta_{23}$
in Eqs. (33) and (38) for mass matrices of types $A_1$ and $A_2$,
respectively. This reciprocity can be, gainfully, used to examine
deviations from maximality for the atmospheric mixing angle for
matrices of types $A_1$ and $A_2$ and, hence, to distinguish
between them.

At $\alpha=90^0$, we have $m_1=s_{12}^2\sqrt{\frac{\Delta m^2}
{\cos 2 \theta_{12}}}$ and $\delta+2\beta=n \pi$ where $n$ is even
for $A_1$ and odd for $A_2$ type neutrino mass matrices. Although,
$\alpha=90^0$ is not allowed, the relation between $\beta$ and
$\delta$ remains approximately valid even for small deviations
about $\alpha=90^0$.

\begin{table}[t]
\begin{center}
\begin{tabular}{|c|ccccc|}
\hline
 Confidence level& $m_1(10^{-3}eV)$& $\alpha(\deg)$ & $\beta(\deg)$ & $\theta_{13}(\deg)$&  $J$  \\
\hline
 $1\sigma$ C.L. &$2.8$ - $4.2$  & $83.4$ - $96.6$  & $-59.4$ - $59.4 $ & $5.6$ - $6.3$ & $-0.011$ - $0.011$ \\
 $2\sigma$ C.L. &$2.0$ - $10.0$ & $76.4$ - $103.6$ & $-90$ - $90 $ & $4.6$ - $9.8$ & $-0.021$ - $0.021$ \\
 $3\sigma$ C.L. &$1.6$ - $16.2$ & $71.2$ - $108.8$ & $-90$ - $90 $ & $3.5$ - $12.4$& $-0.046$ - $0.046$ \\
\hline
\end{tabular}
\end{center}
\caption{The predictions for neutrino mass matrices of class A.}
\end{table}

The numerical analysis for the neutrino mass matrices of type
$A_1$ is done in the following manner. The oscillation parameters
$\Delta m^2_{12}$, $\Delta m^2_{23}$, $s^2_{12}$ and $s^2_{23}$
are varied within their experimental ranges given in Eqs. (11) and
$\alpha$ is calculated as a function of $m_1$ by using Eq. (33).
For these ($m_1$,$\alpha$) values, $\beta$ is calculated from Eqs.
(16) and (17), $\delta$ from Eqs. (29) and (30) and $\theta_{13}$
from Eq. (34). The CHOOZ bound on $\theta_{13}$ is used to
constrain $m_1$ and, hence, all the other neutrino parameters. The
neutrino mass matrices of type $A_2$ can be studied in an
analogous manner. The results have been summarized in Table 1 at
various confidence levels for $m_1$, $\alpha$, $\beta$,
$\theta_{13}$ and Jarlskog rephasing invariant quantity \cite{16}
\begin{equation}
J=s_{12}s_{23}s_{13}c_{12}c_{23}c_{13}^2 \sin \delta.
\end{equation}
These quantities are the same for mass matrices of types $A_1$ and
$A_2$. It can bee seen from Table 1 that the 3 $\sigma$ lower
bound on $\theta_{13}$ is $3.5^0$. The range for the Majorana-type
CP-violating phase $\beta$ at 1 $\sigma$ C.L. is found to be
$-59.4^0$ - $59.4^0$. However, if the neutrino oscillation
parameters are allowed to vary beyond their present 1.2 $\sigma$
C.L. ranges, the full range for $\beta$ ($-90^0$ - $90^0$) is
allowed. As noted earlier, the neutrino mass matrices of type
$A_1$ and $A_2$ differ in their predictions for $\delta$ and
$\theta_{23}$. At one standard deviation, the allowed range of
$\delta$ is ($-110.8^0$ - $110.8^0$) for type $A_1$ and ($69.2^0$
- $290.8^0$) for type $A_2$ and the allowed range of $\theta_{23}$
is ( $44.0^0$ - $48.2^0$) for type $A_1$ and ($41.8^0$ - $46.0^0$)
for type $A_2$ [Table 2]. Just like $\beta$, no constraint on
$\delta$ is obtained above 1.2 $\sigma$ C.L. The neutrino mass
matrices of types $A_1$ and $A_2$ have some overlap in their
predictions regarding $\delta$ and $\theta_{23}$ at one standard
deviation. Therefore, a precise measurement of the oscillation
parameters is necessary to distinguish the neutrino mass matrices
of types $A_1$ and $A_2$.

\begin{table}[b]
\begin{center}
\begin{tabular}{|c|cc|}
\hline

 1 $\sigma$ predictions      &   $\delta$          & $\theta_{23}    $  \\
\hline
 $A_1$ &$-110.8^0$ - $110.8^0$ & $44.0^0$ - $48.2^0$  \\
 $A_2$ &$69.2^0$ - $290.8^0$   & $41.8^0$ - $46.0^0$  \\
\hline
\end{tabular}
\end{center}
\caption{The predictions for $\delta$ and $\theta_{23}$.}
\end{table}

In Fig. 1, we depict the allowed values of $m_1$, $\alpha$ and
$\beta$ as the correlation plots at one standard deviation. The
upper panel shows $\alpha$ as a function of $m_1$ and the lower
panel shows $\beta$ as a function of $\alpha$. The large spread in
the ($m_1$,$\alpha$) plot is due to the errors in the neutrino
oscillation parameters. However, there is a very strong
correlation between $\alpha$ and $\beta$ as indicated by Eqs. (16)
and (17). As seen earlier, the points $\alpha=90^0$ and
$\beta=0^0$ are excluded in Fig. 1. The two Majorana phases are
strongly correlated with each other. Such a correlation between
the Majorana phases was noted earlier \cite{3}. However, full
ranges ($-90^0$ - $90^0$) are allowed for the two Majorana phases
in that analysis \cite{3}. In contrast, we obtain a very narrow
range for the Majorana phase $\alpha$ around $90^0$ [Table 1].
Similarly, the Majorana phase $\beta$ is, also, constrained if the
oscillation parameters are limited to their 1 $\sigma$ ranges. In
Fig. 2, we depict the correlation plots of $\alpha$ and $\beta$
with $\delta$ for matrices of types $A_1$ (left panel) and $A_2$
(right panel). We, also, show the correlation plots of $\delta$
and $\theta_{23}$ with one another as well as with $\theta_{13}$
in Fig. 2. The Dirac-type CP-violating phase $\delta$ is strongly
correlated with the Majorana-type CP-violating phases $\alpha$ and
$\beta$ [Cf. Eqs. (29) and (31)]. It can be seen from
($\alpha$,$\delta$) and ($\beta$,$\delta$) correlation plots that
there are small deviations in the values of $\alpha$ around $90^0$
and the correlation between $\beta$ and $\delta$ is almost linear.
The fact that $\delta+2\beta\simeq 0^0$ for $A_1$ type mass
matrices and $\delta+2\beta\simeq 180^0$ for $A_2$ type mass
matrices is apparent from the ($\beta$,$\delta$) correlation plots
given in the left and right panels in Fig. 2, respectively. The
($\beta$,$\delta$) correlation was noted earlier by Xing \cite{3}
in its approximate form in a different parameterization. The
($\delta$, $\theta_{23}$) plots, clearly, illustrate the point
that the neutrino mass matrices of types $A_1$ and $A_2$ have
different predictions for these variables [Table 2] and only a
limited region is allowed on the ($\delta$, $\theta_{23}$) plane.
This is contrary to the analysis done by Xing \cite{3} where no
constraints on $\delta$ and $\theta_{23}$ have been obtained. In
fact, this feature is crucial for distinguishing mass matrices of
types $A_1$ and $A_2$ which were found to be degenerate in the
earlier analyses \cite{1,2,3}. The constraints on $\delta$ and
$\theta_{23}$ are very sensitive to the values of $\theta_{13}$.
For the values of $\theta_{13}$ smaller than $1 \sigma$ CHOOZ
bound, the constraints on $\delta$ and $\theta_{23}$ become
stronger which can be seen from the ($\theta_{13}$, $\theta_{23}$)
and ($\theta_{13}$,$\delta$) plots. For example, if
$\theta_{13}<6^0$, then $\theta_{23}>46^0$ (above maximal) for
type $A_1$ and $\theta_{23}<44^0$ (below maximal) for type $A_2$.
It can, also, be seen from the ($\theta_{13}$, $\theta_{23}$)
correlation plot in Fig. 2 that the deviation of $\theta_{23}$
from maximality is larger for smaller values of $\theta_{13}$.
Therefore, if future experiments measure $\theta_{13}$ below its
present $1\sigma$ bound, the neutrino mass matrices of types $A_1$
and $A_2$ will have different predictions for $\delta$ and
$\theta_{23}$ with no overlap. However, it would be difficult to
differentiate between matrices of types $A_1$ and $A_2$ if
$\theta_{13}$ is found to be above its present $1\sigma$ range. As
noted earlier, different quadrants for $\delta$ are selected for
neutrino mass matrices of types $A_1$ and $A_2$. It can, also, be
seen in Fig. 2 that the points $\delta=0,\pi$ are ruled out
implying that neutrino mass matrices of class A are necessarily
CP-violating. This feature is, also, apparent in Fig. 3 where we
have plotted $J$ as a function of $m_1$ and in  Fig. 4 (identical
for $A_1$ and $A_2$) which depicts J as a function of $\delta$ for
types $A_1$ (left panel) and $A_2$ (right panel). The symmetry
argument summarized in Eq. (8) is consistent with the numerical
results presented in Fig. 2.

As mentioned earlier, the accuracies of the oscillation parameters
required to distinguish between the theoretical predictions of
neutrino mass matrices of types $A_1$ and $A_2$ depend on the
upper bound on $\theta_{13}$. With the current precision of the
oscillation parameters, the neutrino mass matrices of types $A_1$
and $A_2$ can be distinguished at 1 $\sigma$, 2 $\sigma$ and 3
$\sigma$ C.L. for $\theta_{13}<6.2^o,5.6^o$ and $5.0^o$
respectively. For $\theta_{13}>6.2^o$ the predictions of neutrino
mass matrices of types $A_1$ and $A_2$ can not be differentiated
with the presently available accuracies of the oscillation
parameters. Even with the reduction of the errors in the
measurement of $\Delta m^2_{23}$ to $6\%$ expected at MINOS
\cite{17} and to $2\%$ expected at K2K \cite{18}, no significant
improvement in the prospects for distinguishing between neutrino
mass matrices of types $A_1$ and $A_2$ results [Cf. Table 3]. We
have assumed the central value of atmospheric mixing angle
$\theta_{23}$ to be $45^o$ for the calculations presented in Table
3. If the future experiments measure $\theta_{23}$ to be above or
below maximality at a good statistical significance, the mass
matrices of types $A_1$ and $A_2$ will be distinguishable easily.

\begin{table}[t]
\begin{center}

\begin{tabular}{|c|ccc|}
\hline
     C.L.     &     1 $\sigma$      &     2 $\sigma$      &     3 $\sigma$      \\
     \hline
 Present data & $\theta_{13}<6.2^o$ & $\theta_{13}<5.6^o$ &  $\theta_{13}<5.0^o$  \\
    MINOS     & $\theta_{13}<6.5^o$ &  $\theta_{13}<6.0^o$  & $\theta_{13}<5.5^o$ \\
     K2K      & $\theta_{13}<6.6^o$ & $\theta_{13}<6.2^o$ & $\theta_{13}<5.7^o$ \\
\hline
\end{tabular}
\end{center}
\caption{The upper bounds on $\theta_{13}$ at which the
predictions of neutrino mass matrices of $A_1$ and $A_2$ differ at
various confidence levels for the present and future neutrino
experiments.}
\end{table}

If neutrinoless double beta decay searches give positive results
and $m_{ee}$ is measured experimentally or the atmospheric/reactor
neutrino oscillation experiments confirm inverted hierarchy, the
neutrino mass matrices of class A will be ruled out. A generic
prediction of this class of models is a lower bound on
$\theta_{13}$ [Table 1]. If the forthcoming neutrino oscillation
experiments measure $\theta_{13}$ below $3.5^0$, the neutrino mass
matrices of class A will, again, be ruled out. The forthcoming
neutrino experiments will definitely probe this range of
$\theta_{13}$ and aim at measuring the Dirac type CP violating
phase $\delta$ and the deviations of the atmospheric mixing angle
from maximality \cite{19}. The results of these experiments will
fall in one of the following four categories of phenomenological
interest:
\begin{enumerate}
\item $\theta_{23}>45^0$ and $-90^0<\delta<90^0$,
\item $\theta_{23}<45^0$ and $90^0<\delta<270^0$,
\item $\theta_{23}<45^0$ and $-90^0<\delta<90^0$,
\item $\theta_{23}>45^0$ and $90^0<\delta<270^0$.
\end{enumerate}
If the experiments confirm the first possibility, the neutrino mass matrices
of type $A_1$ may explain the data. If the experiments confirm the
second possibility, the mass matrices of type $A_2$ may be allowed. In
case, the experiments select the third or fourth possibility, the
neutrino mass matrices of class A will be ruled out.

The four cases given above are degenerate for the present neutrino
oscillation data because of the octant degeneracy of $\theta_{23}$
(i.e. if $\theta_{23}<45^0$ or $\theta_{23}>45^0$) and the
intrinsic degeneracy in the sign of $\cos\delta$. The above two
two-fold degeneracies combined with the two-fold degeneracy in the
sign of $\Delta m^2_{23}$ gives rise to the eightfold degeneracy
in the neutrino parameter space which has been studied extensively
in the literature \cite{20,21,22,23}. A specific project named
Tokai-to-Kamioka-Korea (T2KK) two detector complex which will
receive neutrino superbeams from J-PARC \cite{23} has been
proposed to resolve these degeneracies. The intrinsic degeneracy
in the sign of $\cos\delta$ will be resolved by the spectrum
information at T2KK while the $\theta_{23}$ octant degeneracy will
be resolved by observing the difference in $\Delta m^2_{12}$
oscillation effects between the intermediate and far detectors at
T2KK. However, we find that the degeneracies in the neutrino
parameter space are inextricably linked to the degeneracies in the
neutrino mass matrix and by finding the octant of $\theta_{23}$
and sign of $\cos\delta$, the future experiments will be able to
settle the fate of neutrino mass matrices of types $A_1$ and
$A_2$.

If the neutrino mass matrices of class A are confirmed
experimentally, this would correlate the CP-violation induced by
the Dirac phase $\delta$ (which can be observed in the lepton
number conserving processes) with the CP-violation induced by
Majorana phases (which can be observed in the lepton number
violating processes) in a definite way.

In conclusion, the degeneracy between $A_1$ and $A_2$ type
neutrino mass matrices has been examined in detail. In the earlier
analyses \cite{1,2,3}, textures within a class were found to be
experimentally indistinguishable. However, our analysis shows that
the knowledge of the deviations of the atmospheric mixing from the
maximality and of the quadrant of the Dirac-type CP-violating
phase $\delta$ can be used to lift this degeneracy between mass
matrices of types $A_1$ and $A_2$. Contrary to the results of
earlier analyses \cite{3}, which leave the Dirac phase completely
unconstrained, we have obtained interesting constraints not only
for the Dirac phase but, also, for the Majorana-type CP-violating
phases. Our analysis, also, shows that the prospects for the
measurement of $\theta_{13}$ for neutrino mass matrices of class A
are quite optimistic. Moreover, the neutrino mass matrices of
class A connect the deviation of $\theta_{23}$ from maximality
with the quadrant of $\delta$. Similarly, the CP-violation
observed in the neutrino oscillation experiments is linked with
the CP-violation induced by Majorana phases.

\vspace{30pt}
\textit{\large{Acknowledgments}}

The research work of S. D. and S. V. is supported by the Board
of Research in Nuclear Sciences (BRNS), Department of Atomic
Energy, Government of India \textit{vide} Grant No. 2004/ 37/ 23/
BRNS/ 399. S. K. acknowledges the financial support provided by
Council for Scientific and Industrial Research (CSIR), Government
of India. We would like to thank Manmohan Gupta for critical
reading of the manuscript and helpful suggestions.

\pagebreak

\newpage

%%%%%%%%%%%%%%%%%%%%%%%%%%%%%%%%%%%%%%%%%%%%%%%%%%%%%%%%%%%%%%%%%%

\begin{figure}[tb]
\begin{center}
%\vskip 1cm
{\epsfig{file=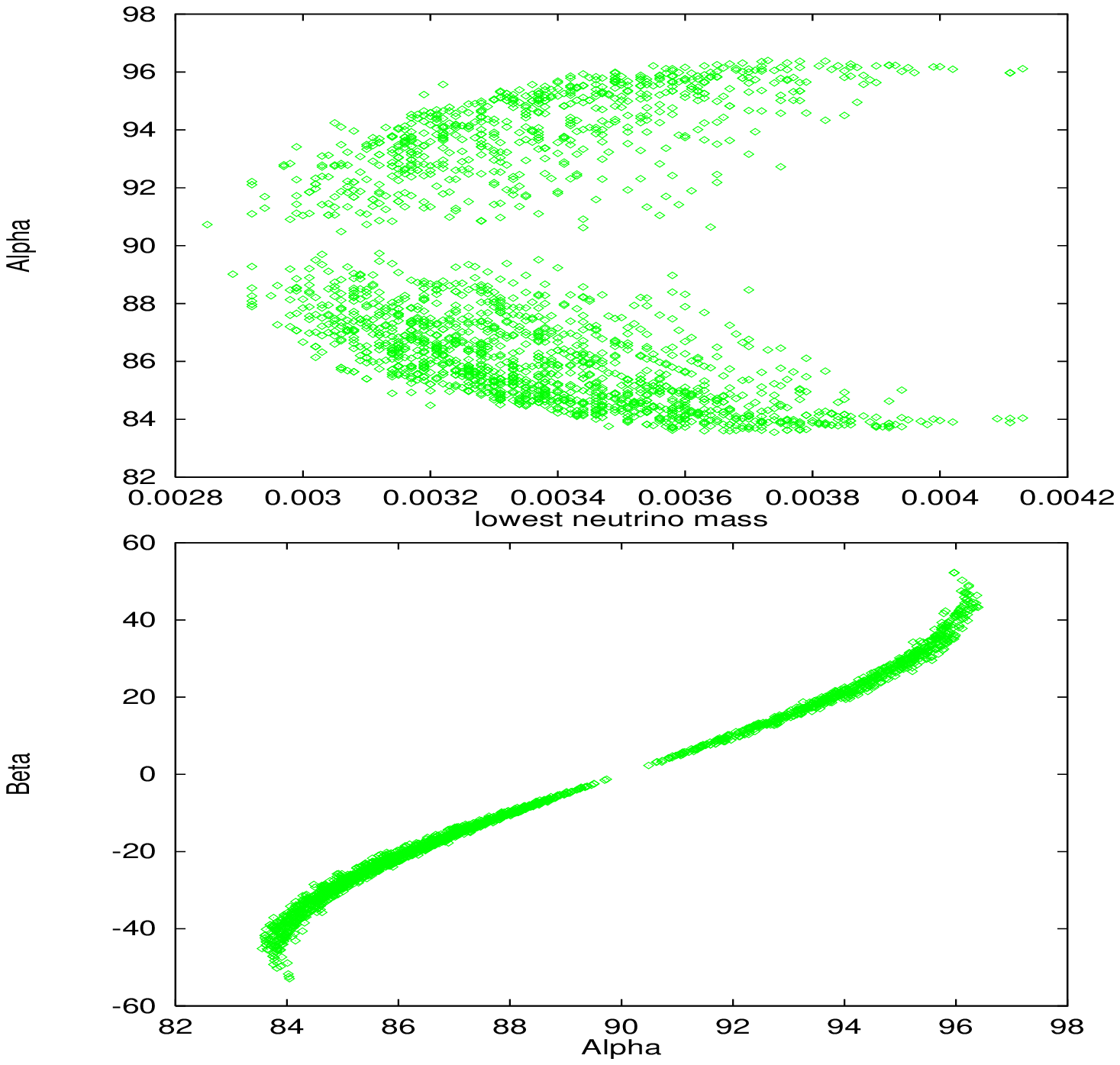, width=10cm,height=10cm}}
\end{center}
\caption{Correlation plots for neutrino mass matrices of class A
at one standard deviation.}
\end{figure}
%%%%%%%%%%%%%%%%%%%%%%%%%%%%%%%%%%%%%%%%%%%%%%%%%%%%%%%%%%%%%%%%%%%%

%%%%%%%%%%%%%%%%%%%%%%%%%%%%%%%%%%%%%%%%%%%%%%%%%%%%%%%%%%%%%%%%%%

\begin{figure}[tb]
\begin{center}
%\vskip 1cm
{\epsfig{file=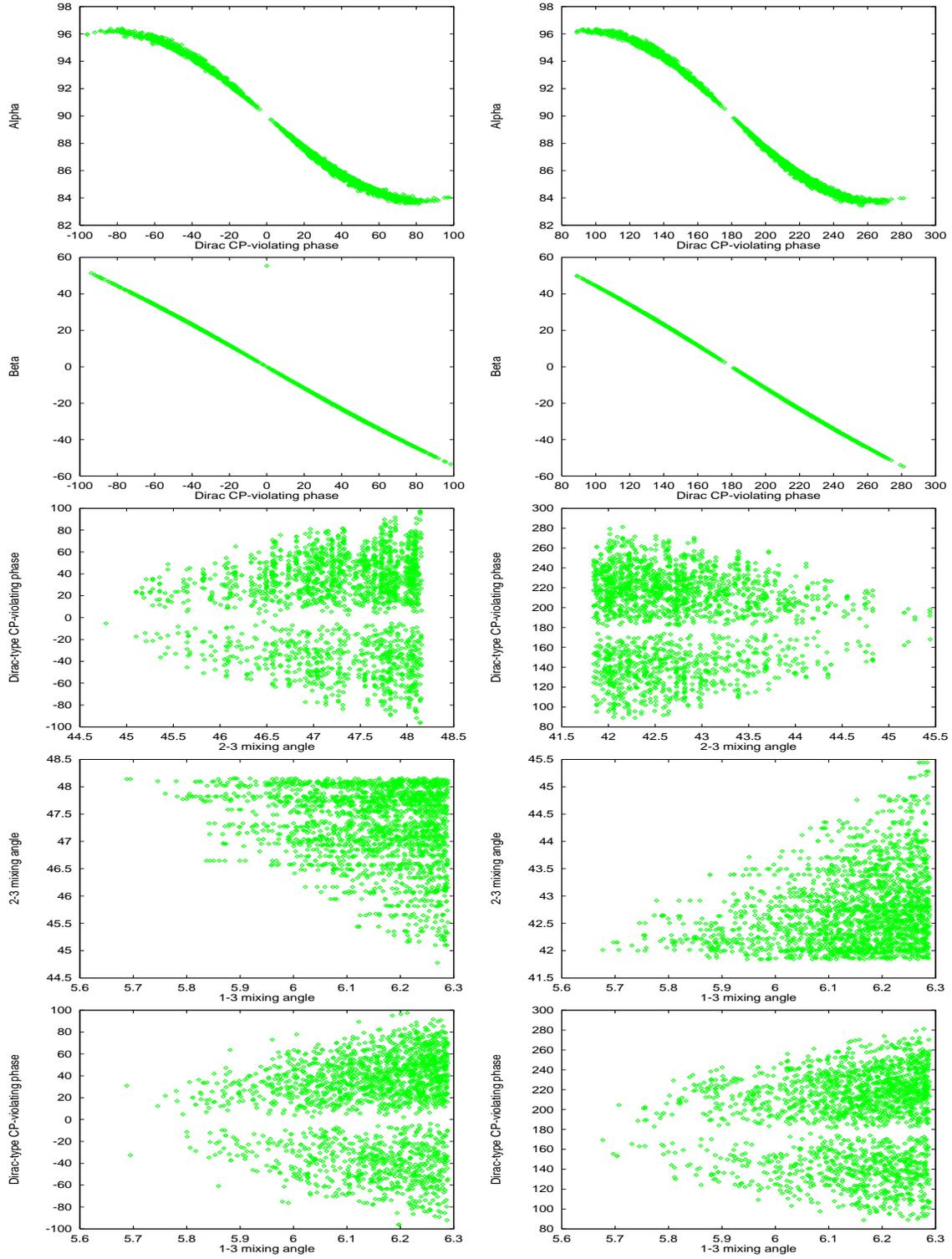, width=15cm,height=20cm}}
\end{center}
\caption{Correlation plots for neutrino mass matrices of types
$A_1$ (left panel) and $A_2$ (right panel).}
\end{figure}
%%%%%%%%%%%%%%%%%%%%%%%%%%%%%%%%%%%%%%%%%%%%%%%%%%%%%%%%%%%%%%%%%%%

%%%%%%%%%%%%%%%%%%%%%%%%%%%%%%%%%%%%%%%%%%%%%%%%%%%%%%%%%%%%%%%%%%

\begin{figure}[t]
\begin{center}
%\vskip 1cm
{\epsfig{file=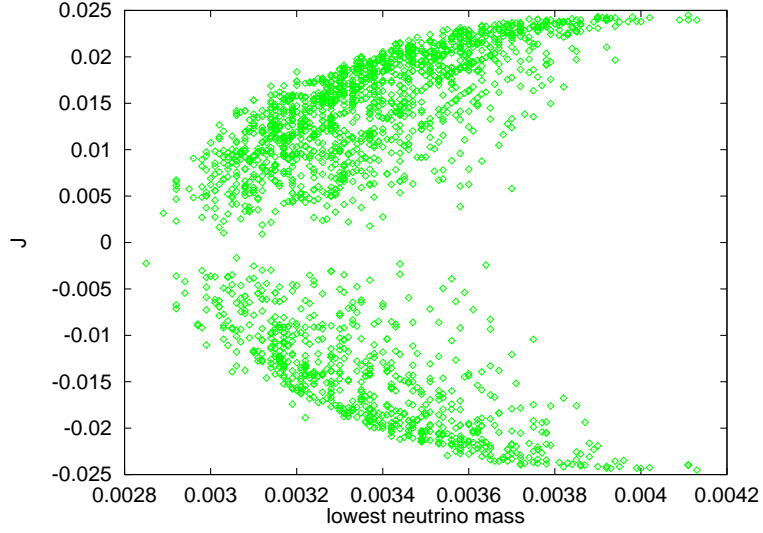, width=10cm,height=7cm}}
\end{center}
\caption{$J$ as a function of $m_1$ for the neutrino mass matrices
of class A.}
\end{figure}
%%%%%%%%%%%%%%%%%%%%%%%%%%%%%%%%%%%%%%%%%%%%%%%%%%%%%%%%%%%%%%%%%%%%

%%%%%%%%%%%%%%%%%%%%%%%%%%%%%%%%%%%%%%%%%%%%%%%%%%%%%%%%%%%%%%%%%%

\begin{figure}[t]
\begin{center}
%\vskip 1cm
{\epsfig{file=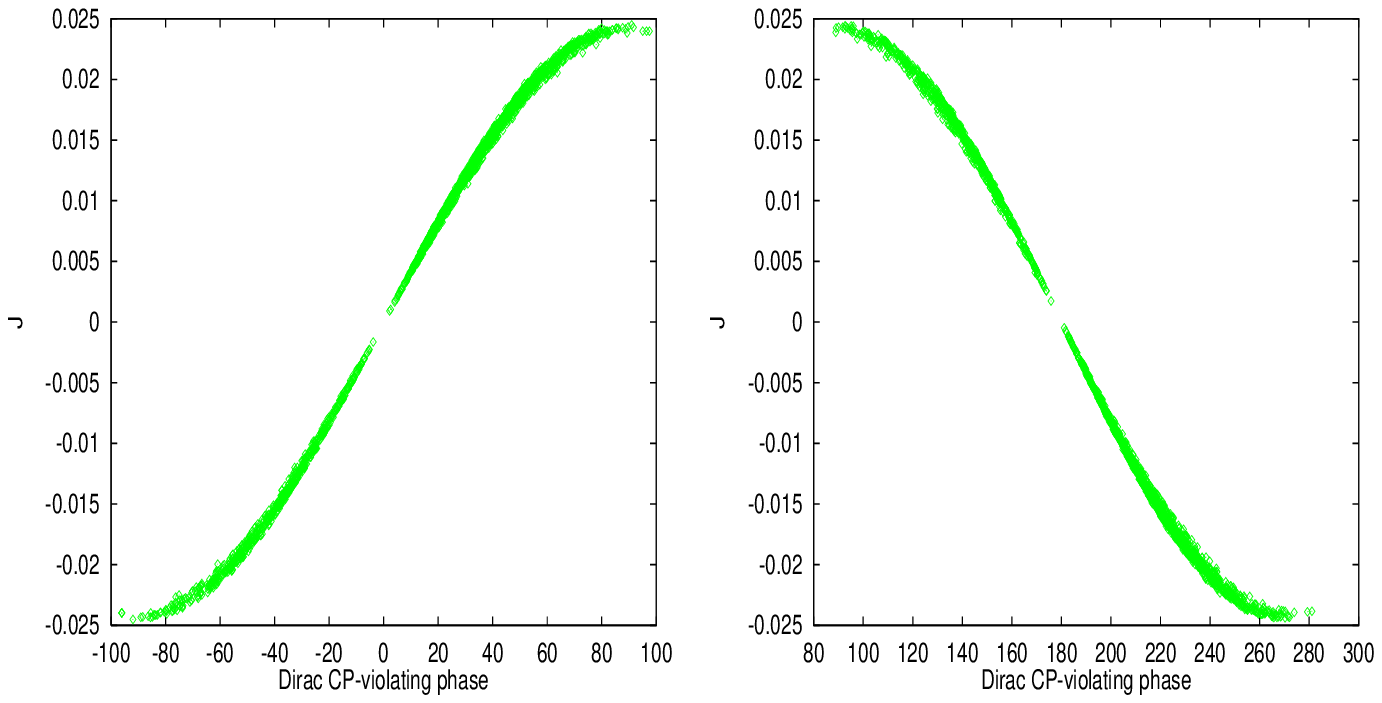, width=15cm,height=7cm}}
\end{center}
\caption{$J$ as a function of $\delta$ for the neutrino mass matrices
of types $A_1$ (left panel) and $A_2$ (right panel).}
\end{figure}
%%%%%%%%%%%%%%%%%%%%%%%%%%%%%%%%%%%%%%%%%%%%%%%%%%%%%%%%%%%%%%%%%%%%


\begin{thebibliography}{99}

\bibitem{1} Paul H. Frampton, Sheldon L. Glashow and Danny
Marfatia, \textit{Phys. Lett.} \textbf{B 536}, 79 (2002).

\bibitem{2} Bipin R. Desai, D. P. Roy and Alexander R. Vaucher,
\textit{Mod. Phys. Lett} \textbf{A 18}, 1355 (2003).

\bibitem{3} Zhi-zhong Xing, \textit{Phys. Lett.} \textbf{B 530}
159 (2002); Wanlei Guo and Zhi-zhong Xing, \textit{Phys. Rev.}
\textbf{D 67}, 053002 (2003).

\bibitem{4} Alexander Merle and Werner Rodejohann, \textit{Phys. Rev.}
\textbf{D 73}, 073012 (2006).

\bibitem{5} G. C. Branco, R. Gonzalez Felipe, F. R. Joaquim and T. Yanagida,
Phys. Lett. \textbf{B 562} 265 (2003).

\bibitem{6} Bhag C. Chauhan, Joao Pulido and Marco Picariello,
\textit{Phys. Rev.} \textbf{D 73}, 053003 (2006).

\bibitem{7} Xiao-Gang He and A. Zee, \textit{Phys. Rev.} \textbf{D 68}, 037302 (2003).

\bibitem{8} Suraj. N. Gupta and Subhash Rajpoot, ``Quark Mass Matrices and the Top
Quark Mass'', Wayne State University Preprint, September 1990; S.
N. Gupta and J. M. Johnson, \textit{Phys. Rev.} \textbf{D 44},
2110 (1991); S. Rajpoot, \textit{Mod. Phys. Lett.} \textbf{A 7},
309 (1992); H. Fritzsch and Z.Z. Xing, \textit{Phys. Lett.}
\textbf{B 353}, 114 (1995).

\bibitem{9} H. C. Goh, R. N. Mohapatra and
Siew-Phang Ng, \textit{Phys. Rev.} \textbf{D 68}, 115008 (2003).


\bibitem{10} P. H. Frampton, M. C. Oh and T. Yoshikawa,\textit{Phys. Rev.}
\textbf{D 66}, 033007 (2002).

\bibitem{11} Walter Grimus and Luis Lavoura, \textit{J. Phys.}
\textbf{G 31}, 693-702 (2005).

\bibitem{12} S. Kaneko, M. Katsumata and M. Tanimoto, \textit{JHEP}
\textbf{0307}, 025 (2003).

\bibitem{13}  G. L. Fogli \textit{et al}, \textit{Prog. Part. Nucl. Phys.}
\textbf{57}, 742-795 (2006).
\bibitem{14}M. Maltony, T. Schwetz, M. A. Tortola and J. W. F.
Valle,\textit{ New J. Phys.} \textbf{6}, 122 (2004).

\bibitem{15} S. Dev and Sanjeev Kumar, \textit{Mod. Phys. Lett.} \textbf{A} (in press);
hep-ph/0607048.

\bibitem{16} C. Jarlskog \textit{Phys. Rev. Lett.} \textbf{55}, 1039
(1985).

\bibitem{17} D. G. Michael et al. [MINOS Collaboration],
\textit{Phys. Rev. Lett.} \textbf{97}, 191801 (2006).

\bibitem{18} M. H. Ahn et al. [K2K Collaboration], \textit{Phys. Rev.}
\textbf{D 74}, 072003 (2006); hep-ex/0606032.

\bibitem{19} E. Abouzaid \textit{et al}, Report of the APS
Neutrino Study Reactor Working Group, 2004.

\bibitem{20} V. Barger, D. Marfatia and K. Whisnant,  \textit{Phys. Rev.}
\textbf{D 65}, 073023 (2002).
\bibitem{21} Hisakazu Minakata, Hiroshi Nunokawa, Stephen Parke,
 \textit{Phys. Rev.}
\textbf{D 66}, 093012 (2002).

\bibitem{22} Takaaki Kajita, Hisakazu Minakata, Shoei Nakayama, and Hiroshi
Nunokawa,  \textit{Phys. Rev.} \textbf{D 75}, 013006 (2007).

\bibitem{23} Hisakazu Minakata, hep-ph/0701070.

\end{thebibliography}
\end{document}